\documentclass[aps,prc,preprint,amsmath,amssymb,showpacs,preprintnumbers,superscriptaddress]{revtex4-1}
\usepackage{CJK}
\usepackage{graphicx}
\usepackage{dcolumn}
\usepackage{bm}
\usepackage{mathrsfs}   
\usepackage{color,xcolor}  
\usepackage{multirow}  
\usepackage{booktabs} 

\usepackage{hyperref}

\newcommand{\bbm}{\begin{bmatrix}}
\newcommand{\ebm}{\end{bmatrix}}
\newcommand{\bBm}{\begin{Bmatrix}}
\newcommand{\eBm}{\end{Bmatrix}}
\newcommand{\bpm}{\begin{pmatrix}}
\newcommand{\epm}{\end{pmatrix}}

\begin{document}


\title{In-medium nucleon-nucleon cross sections from relativistic \textit{ab initio} calculations}

\author{Tianyu Wang}
\affiliation{School of Physics and Mechatronic Engineering, Guizhou Minzu University, Guiyang 550025, China}

\author{Hui Tong}
\affiliation{Helmholtz-Institut f\"ur Strahlen- und Kernphysik and Bethe Center for Theoretical Physics, Universit\"at Bonn, Bonn D-53115, Germany}

\author{Chencan Wang}
\affiliation{Graduate School of China Academy of Engineering Physics, Beijing 100193, China}
\author{Xiaoying Qu}
\email{Corresponding author: xiaoyingqu@126.com}
\affiliation{School of Physics and Mechatronic Engineering, Guizhou Minzu University, Guiyang 550025, China}

\author{Sibo Wang}
\email{Corresponding author: sbwang@cqu.edu.cn}
\affiliation{Department of Physics and Chongqing Key Laboratory for Strongly Coupled Physics, Chongqing University, Chongqing 401331, China}

\date{\today}

\begin{abstract}

The in-medium nucleon-nucleon scattering cross section is a pivotal quantity for studying the medium effects of strong interaction, and its precise knowledge is critical for understanding the equation of state for dense matter, intermediate-energy heavy-ion collision dynamics, and related phenomena. In this work, we perform a microscopic investigation of in-medium nucleon-nucleon scattering cross sections, by utilizing the relativistic Brueckner-Hartree-Fock (RBHF) theory with the Bonn potential. The fully incorporation of both positive- and negative-energy states in the RBHF solutions allows us to determine the single-particle potentials, the effective $G$ matrix, and the scattering cross section uniquely. The momentum, density, and isospin dependence of the cross section for $pp$, $nn$, and $np$ scattering are studied in detail. Our results provide a solid foundation for future parameterization studies of multi-parameter dependency of total scattering cross sections.

\end{abstract}



\maketitle


\section{Introduction}\label{SecI}

Since Yukawa's pioneering proposal of the meson-exchange theory in 1935 \cite{1935-Yukawa-PPMSJ}, the study of strong interactions has undergone more than nine decades of extensive development. Despite numerous groundbreaking achievements in both theoretical understanding and experimental investigations, the structural and dynamical properties of strongly interacting systems, particularly those under extreme conditions of density, temperature, and isospin asymmetry, remain at the forefront of contemporary research. These fundamental questions continue to attract significant attention across multiple disciplines, including particle physics, nuclear physics, and astrophysics.

The in-medium nucleon-nucleon (NN) scattering cross sections represent fundamental physical quantities in the study of medium effects of the strong interaction. First, as the lightest strongly interacting fermions, nucleons and their bound states constitute the predominant form of visible matter in our universe. Second, NN interactions are crucial in the low-energy regime of the strong interaction, directly connected to the non-perturbative aspects of quantum chromodynamics (QCD) \cite{Ishii-2007-Phys.Rev.Lett.99.022001}. Finally, the in-medium NN scattering cross sections govern the mean free path of nucleons \cite{2008-Sammarruca-PhysRevC.77.047301}, thereby determining the transparency and other essential transport properties of nuclear matter.

In the realm of neutron star physics, the in-medium NN scattering cross sections play a pivotal role in determining fundamental astrophysical properties through their influence on effective NN interactions, including the equation of state, mass-radius relation, and tidal deformabilities \cite{Burgio-2021_PPNP120.103879}. In addition to these static characteristics, the in-medium NN scattering cross sections significantly impact the dynamical evolution of neutron stars by modulating key thermal properties, such as thermal conductivity, shear viscosity, and neutrino emissivity \cite{2004-Yakovlev-ASR}. Furthermore, extending the scope to include strange baryons, the in-medium baryon-baryon scattering cross sections involving hyperons are intimately related to the longstanding problem "hyperon puzzle" \cite{2023-Sedrakian-PPNP, TongH2025SB, Tong:2025sui}, which remains one of the most challenging issues in contemporary nuclear astrophysics.

In the context of intermediate-energy heavy-ion collisions (HICs), the in-medium NN scattering cross sections serve as crucial fundamental parameters for theoretical frameworks and transport model simulations \cite{Xu2019PPNP106, 2020-ZhangYX-FoP, 2022-Wolter-PPNP}. At the microscopic level, these cross sections govern both the collision rates and angular dependence in NN collisions, thus directly influencing the production mechanisms, dynamical evolution, and spatial distribution of emitted particles, as well as the emergence and characteristics of collective flow. Experimentally, the measured final-state particle multiplicities, isotopic compositions, and kinematic distributions exhibit significant sensitivity to the in-medium modifications of NN interactions. This dependence enables the extraction of essential nuclear matter properties through systematic comparisons between theoretical predictions and experimental measurements in HICs.

The accurate determination of in-medium NN scattering cross sections presents significant theoretical challenges \cite{Cugnon1996, 2022-HanSC-PhysRevC.106.064332}, primarily stemming from their complex dependence on various parameters. These include not only the intrinsic properties of the colliding nucleon pair, such as relative momentum, center-of-mass motion, and isospin projections, but also the environmental conditions characterized by baryon density, temperature, and isospin asymmetry. Current theoretical studies typically incorporate medium modifications through either empirical density-dependent scaling factors applied to vacuum cross sections \cite{2001-Liu-PhysRevLett.86.975, 2012-ZhangYX-PhysRevC.85.024602, 2011-LiQF-PhysRevC.83.044617, 2020-WangR-PLB} or the use of effective nucleon masses in the nuclear medium \cite{Pandharipande1992_PRC45-791, 2002-Persram-PhysRevC.65.064611, LiBA2005_PRC72-064611}. While these phenomenological approaches provide valuable connections to experimental data, they fall short of capturing the fundamental nature of strongly interacting systems. A more comprehensive understanding requires \textit{ab initio} calculations that simultaneously combine realistic NN interactions with advanced many-body techniques \cite{2023-Machleidt-FBS, 2024-Machleidt-PPNP}.

The realistic NN interactions, which accurately reproduce free-space NN scattering data and two- and three-body bound states, can be systematically classified into three main categories based on their theoretical foundations. The first category comprises potential models constructed through operator analysis, exemplified by the Argonne V18 potential \cite{Wiringa-1995-PhysRevC.51.38}. The second category utilizes meson-exchange models, such as the Bonn potential \cite{Machleidt1989_ANP19-189} and its refined version, the CD-Bonn potential \cite{2001-Machleidt-PhysRevC.63.024001}. The third category encompasses chiral interactions derived from chiral effective field theory, which have gained increasing attention in recent years \cite{2003-Entem-PhysRevC.68.041001, 2022-LuJX-PhysRevLett.128.142002}. Comprehensive reviews and detailed discussions of these interaction models can be found in several authoritative references \cite{Machleidt1989_ANP19-189, Machleidt2011, 2009-Epelbaum-RevModPhys.81.1773, 2024-Machleidt-PPNP}.

Based on realistic NN interactions, various microscopic many-body approaches have been developed to investigate in-medium NN elastic and inelastic scattering cross sections. These include the Brueckner-Hartree-Fock (BHF) theory \cite{Schulze1997_PRC55-3006, 2020-ShangXL-PhysRevC.101.065801}, the Green's function approach \cite{Alm1994_PRC50-31,Alm1995}, the variational method \cite{Pandharipande1992_PRC45-791}, the relativistic Boltzmann-Uehling-Uhlenbeck approach \cite{1994-MaoGJ-PhysRevC.49.3137, 2017-LiQF-PLB}, and the relativistic Brueckner-Hartree-Fock (RBHF) theory \cite{GQLi1993_PRC48-1702,CFuchs2001_PRC64-024003,2006-Sammarruca-PhysRevC.73.014001,2014-Sammarruca-EPJA}. General agreement on several key features has been reached: the reduction of cross sections for low energy scattering in the nuclear medium, a strong enhancement near the Fermi surface, and the isospin symmetry breaking to some extent. However, quantitative predictions vary depending on the extent of medium effects incorporated in each theoretical framework. The RBHF theory distinguishes itself by consistently accounting for multiple crucial effects: Pauli exclusion principle, energy-momentum dispersion relations, and relativistic dynamics \cite{Day1967RMP, Anastasio1980_PRL45-2096, SHEN-SH2019_PPNP109-103713, 2020-TongH-PhysRevC.101.035802}. This comprehensive treatment enables RBHF theory to successfully reproduce nuclear matter saturation properties \cite{Brockmann1990_PRC42-1965, WANG-SB2021_PRC103-054319}, a longstanding benchmark for nuclear many-body theories. The superior performance of the RBHF theory as compared to the non-relativistic BHF method is generally attributed to its inherent inclusion of a class of density-dependent three-nucleon forces (3NFs) through virtual nucleon–antinucleon pair excitations, commonly known as the $Z$-diagram \cite{Brown-1987-Phys.Scr.}. This should not be taken as evidence that the relativistic effect is a replacement for 3NFs. Both mechanisms are repulsive in nature, but they represent distinct physical effects that must be treated consistently in a complete theory.

The RBHF theory is fundamentally characterized by its double iterative structure involving the effective interaction $G$ matrix and single-particle potentials \cite{2010-VanGiai-JPG}. Historically, RBHF implementations were restricted to positive-energy solutions of the Dirac equation, with the negative-energy state (NES) contributions incorporated through approximate treatments. This limitation hindered the rigorous determination of single-particle potentials from the $G$ matrix. Recent computational advances have overcome this constraint by implementing fully self-consistent RBHF calculations in the full Dirac space, explicitly including both positive- and negative-energy states \cite{WANG-SB2021_PRC103-054319,SBWang2022_PRC106-L021305,Tong:2022tlt}. This comprehensive approach not only provides unambiguous determination of nucleon single-particle properties but also significantly enhances the precision of $G$ matrix calculations through iterative solutions of the in-medium NN scattering equation. These developments offer a rigorous framework for the theoretical investigation of in-medium NN scattering cross sections.

In this work, we aim to derive the in-medium NN scattering cross sections with the RBHF theory in the full Dirac space. It should be emphasized that one of the largest sources of uncertainty in the determination of in-medium cross sections is the dependence on the underlying NN interaction and 3NFs \cite{2020-Shternin-PhysRevD.102.063010}. In the present study we restrict ourselves to the Bonn A potential \cite{Brockmann1990_PRC42-1965}, since RBHF calculations based on this interaction are known to reproduce the empirical saturation properties of nuclear matter \cite{WANG-SB2021_PRC103-054319}. A systematic comparison with other interactions will be pursued in future work.

This paper is organized as follows: Section~\ref{SecII} describes briefly the theoretical formalism, including the RBHF implementation in the full Dirac space and the methodology for calculating in-medium NN scattering observables. Section~\ref{SecIII} presents and analyzes the numerical results, focusing on medium-modified phase shifts, angular distributions, and integrated cross sections. Section~\ref{SecIV} summarizes the key findings and discusses potential future developments in this research direction.

\section{Theoretical framework} \label{SecII}

\subsection{The RBHF theory in the full Dirac space}

The RBHF theory can be decomposed at multiple levels. At the one-body level, it solves the single-particle Dirac equation, describing the motion of nucleons in the nuclear medium:
\begin{equation}\label{eq:DiracEquation}
  \left[ \bm{\alpha}\cdot\bm{p}+\beta \left(M+\mathcal{U}\right) \right] u(\bm{p},s)
  = E_{\bm{p}}u(\bm{p},s),
\end{equation}
where $\bm{\alpha}$ and $\beta$ are the Dirac matrices, $M$ is the nucleon mass, $\bm{p}$ and $E_{\bm{p}}$ are the momentum and single-particle energy, and $s$ denotes the spin. The single-particle potential $\mathcal{U}$ incorporates the in-medium effects. At the two-body level, RBHF theory solves the in-medium Thompson equation~\cite{Brockmann1990_PRC42-1965} to obtain the effective $G$ matrix:
\begin{equation}\label{eq:ThomEqu}
    \begin{split}
  G(\bm{q}',\bm{q}|\bm{P},W)
  =&\ V(\bm{q}',\bm{q}|\bm{P})
  + \int \frac{d^3k}{(2\pi)^3}
  V(\bm{q}',\bm{k}|\bm{P}) \\
    & \times \frac{M^{*}_{\bm{P}+\bm{k}}M^{*}_{\bm{P}-\bm{k}}}{E^*_{\bm{P}+\bm{k}}E^*_{ \bm{P}-\bm{k}}}
    \frac{Q(\bm{k},\bm{P})}{W-E_{\bm{P}+\bm{k}}-E_{\bm{P}-\bm{k}} + i\epsilon}  G(\bm{k},\bm{q}|\bm{P},W),
  \end{split}
\end{equation}
where $\bm{P} = \frac{1}{2}(\bm{k}_1 + \bm{k}_2)$ is the total center-of-mass momentum and $\bm{k} = \frac{1}{2}(\bm{k}_1 - \bm{k}_2)$ is the relative momentum of the two interacting nucleons with momenta $\bm{k}_1$ and $\bm{k}_2$. The initial, intermediate, and final relative momenta in the two-nucleon scattering process are denoted by $\bm{q}$, $\bm{k}$, and $\bm{q}'$, respectively. The Pauli operator $Q(\bm{k},\bm{P})$ prevents scattering into occupied intermediate states. The starting energy $W$ is taken as the sum of the single-particle energies in the initial state.


At the many-body level, RBHF theory corresponds to the hole-line expansion truncated at the two-hole-line level. Within this approximation, the binding energy of the nuclear many-body system is calculated using the direct and exchange contributions of the $G$ matrix.

For the implementation of RBHF calculations in the full Dirac space, both positive- and negative-energy states of the Dirac equation are considered. This enables the construction of all matrix elements of the single-particle potential operator $\mathcal{U}$ by summing the effective two-body $G$ matrix over all nucleons within the Fermi sea in the full Dirac space. This construction allows for the exact extraction of the Lorentz structure of\  $\mathcal{U}$, which is crucial for solving the Dirac equation in the next iteration. Inclusion of the NESs effectively circumvents approximations inherent in calculations restricted to the positive-energy Dirac space. More theoretical and numerical details are provided in Refs.~\cite{WANG-SB2021_PRC103-054319,SBWang2022_PRC106-L021305,WANG-SB2022-PhysRevC.105.054309}.

\subsection{Calculation of in-medium cross sections}

Starting from the $S$ matrix calculated from the $G$ matrix, the scattering amplitudes $\mathcal{M}$ are constructed by summing all the relevant partial-wave channels \cite{Reinert2017SemilocalMR}
\begin{equation}\label{eq-S2M}
  \begin{split}
    \mathcal{M}^{S}_{m'_S, m_S} = &\ \frac{1}{iq} \sum_{JL'L} C(L',S,J;m_S-m'_S,m'_S,m_S) Y_{L'(m'_S-m_S)} (\hat{\bm{q}})
       (S^{JS}_{L'L} - \delta_{L'L}) \\
       &\times C(L,S,J;0,m_S,m_S) \sqrt{\pi (2L+1)}\ i^{L-L'} [1-(-1)^{L+S+T}].
  \end{split}
\end{equation}
Here, $J$, $S$, and $L', L$ denote the total angular momentum, total spin, and orbital angular momenta, respectively; $T$ is the total isospin of the scattering nucleons; $C(j_1, j_2, j_3;m_1, m_2, m_3)$ are Clebsch-Gordan coefficients; and $Y_{lm}(\hat{\bm{q}})$ are spherical harmonics. In our calculations, all panels-wave $S$ matrices with total angular momentum $J\leq 10\ \hbar$ are summed up.

The scattering amplitudes $\mathcal{M}^{S}_{m'_S, m_S}$ can also be given in matrix form as
\begin{equation}
    \mathcal{M} = \begin{pmatrix}
        \mathcal{M}_{ss} & 0 & 0 & 0 \\
        0 & \mathcal{M}_{--} & \mathcal{M}_{-0} & \mathcal{M}_{-+} \\
        0 & \mathcal{M}_{0-} & \mathcal{M}_{00} & \mathcal{M}_{0+} \\
        0 & \mathcal{M}_{+-} & \mathcal{M}_{+0} & \mathcal{M}_{++}
    \end{pmatrix},
\end{equation}
where $\mathcal{M}_{ss}\equiv \mathcal{M}^{0}_{0,0}$ stands for the spin-singlet $(S=0)$ states, while the subscripts $\pm, 0$ are used for $m_S=\pm1, 0$ in the spin-triplet $(S=1)$ states. Then the unpolarized differential cross section and total cross section are calculated by \cite{JBystricky1978_jphys39}
\begin{equation}
  d\sigma/d\Omega = \frac{1}{4} \mathrm{Tr} \left(\mathcal{M}^\dag \mathcal{M}\right),\qquad
  \sigma(E_\text{lab.}, P, \rho, \alpha) = \int d\Omega\ \frac{d\sigma}{d\Omega} (E_\text{lab.}, P, \rho, \alpha, \theta_{\text{c.m.}}) .
\end{equation}
In this work we don't consider the Pauli blocking of the final state, which would lead to quantitatively and qualitatively different cross sections \cite{2006-Sammarruca-PhysRevC.73.014001, 2014-Sammarruca-EPJA}. However, we notice that the Pauli blocking factor for the final states is always included in the transport models describing nucleus-nucleus collisions.

It is important to emphasize that the in-medium $S$ matrix differs from its free-space counterpart because the $G$ matrix in Eq.~\eqref{eq:ThomEqu} already incorporates the essential medium effects. First, Pauli blocking is enforced by the operator $Q(\bm{k},\bm{P})$, which prevents scattering into occupied intermediate states. Second, dispersion effects are included through the use of effective single-particle energies in the energy denominator, ensuring a consistent treatment of the in-medium dispersion relation. Third, the interaction matrix elements are constructed with effective Dirac spinors, which embed the density dependence of the nucleon wave functions and encode relativistic medium modifications. These three ingredients—Pauli blocking, dispersion, and relativistic dressing—are directly responsible for the density and momentum dependence of the in-medium NN cross sections derived in this work.

\section{Results and discussion}\label{SecIII}

In this section, we present the numerical results obtained using the RBHF theory with the Bonn A potential, which accurately describes the nuclear matter properties at the saturation point. These results include medium-modified phase shifts, angular distributions, and integrated cross sections.

\subsection{Phase shifts in nuclear medium}\label{SubSecI}

\begin{figure}[htbp]
  \makebox[\textwidth][c]{\includegraphics[width=1.0\textwidth]{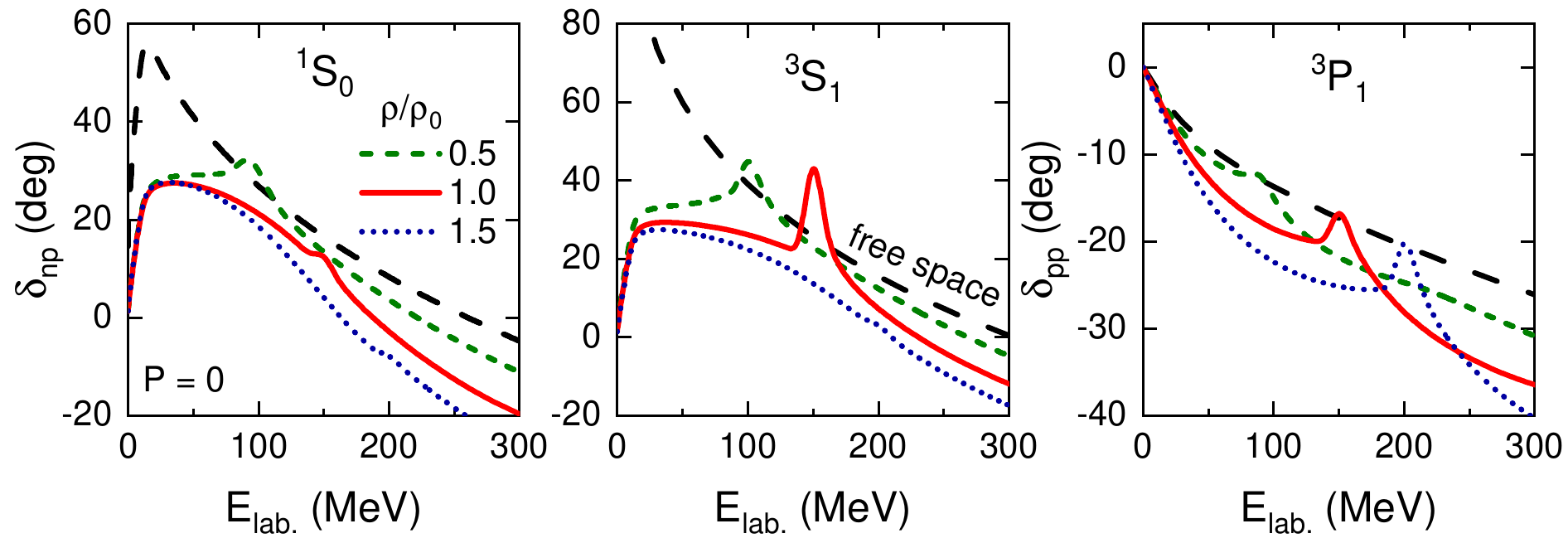}}
  \caption{The $np$ scattering phase shifts for the $^{1}S_{0}$ and $^{3}S_{1}$ channels, and $pp$ scattering phase shifts for the $^{3}P_{1}$ channel as functions of the laboratory energy. In each panel, the green short-dashed, red solid, and blue short-dotted lines represent the results at densities $\rho=0.5\rho_0, \rho_0$, and $1.5\rho_0$, respectively. For comparison, the phase shifts in free space are shown as black dashed lines. The total momentum and isospin asymmetry are set as $P=0$ and $\alpha=0$.}
  \label{Fig1}
\end{figure}

Our numerical codes are first validated by reproducing phase shifts in free space, where the $S$ matrix is unitary. In nuclear medium, the unitarity of $S$ matrix is slightly violated, mainly due to the Pauli blocking. Nonetheless, the in-medium phase shifts are still meaningful since they are irrelevant to the modulus of $S$ matrix. In addition, the derived in-medium phase shifts can serve as valuable pseudo-data for constraining density-dependent terms in energy functionals. These quantities have been used, for example, to constrain the tensor-force strength in energy functionals by fitting to the evolution of spin-orbit splittings in neutron drops obtained from RBHF calculations using the Bonn potential \cite{2018-ShenSH-PLB-ND, 2019-ShenSH-PhysRevC.99.034322, 2019-WangSB-PhysRevC.100.064319}.

The first two panels in Fig.~\ref{Fig1} display the in-medium neutron-proton ($np$) phase shifts for the $^{1}S_{0}$ and $^{3}S_{1}$ channels as functions of the laboratory energy $E_\text{lab.}$ up to the pion-production threshold ($\approx 300$\ MeV). The results are presented for three distinct baryon densities: $\rho = 0.5\rho_0, \rho_0$, and $1.5\rho_0$, where the saturation density $\rho_0$ is $0.16\ \text{fm}^{-3}$. The total momentum and isospin asymmetry are set to $P=0$ and $\alpha=0$, respectively. These two specific channels are selected to examine the medium effects on potential bound states, considering the virtual state in the $^{1}S_{0}$ channel and a true bound state, the deuteron, in the $^{3}SD_{1}$ mixed channel. As illustrated in Figure~\ref{Fig1}, the in-medium phase shifts demonstrate distinctive characteristics: they approach zero at $E_\text{lab.}=0$, indicating the absence of bound states in these channels in nuclear matter.

A notable feature is the sharp enhancement followed by suppression in the $^{3}S_{1}$ phase shift at low laboratory energies, which becomes more prominent at lower densities, indicating a gradual convergence toward the free-space behavior. This density dependence trend also highlights the difficulties encountered in RBHF calculations for densities below $\rho = 0.08\ \text{fm}^{-3}$, where the emergence of nucleon bound states leads to inhomogeneities in the nuclear medium.

Furthermore, a significant enhancement of the phase shift is observed at laboratory energies near the Fermi momentum, most pronounced in the $^{3}S_{1}$ channel and $\rho=\rho_0$. For symmetric nuclear matter at $\rho = 0.16\ \text{fm}^{-3}$, the nucleon Fermi momentum is $k_\text{F} = 1.33\ \text{fm}^{-1}$, corresponding to a laboratory energy of $E_\text{lab.}\approx 147$\ MeV. In the middle panel of Fig.~\ref{Fig1}, the enhancement is observed around 150 MeV. This enhancement indicates a stronger interaction between nucleons at these energies.

To investigate whether the observed enhancement is specific to certain channels, the right panel in Fig.~\ref{Fig1} presents the $pp$ phase shifts for the $^{3}P_{1}$ channel, which does not exhibit virtual or bound states in free space. It is important to note that only the phase shifts arising from the strong interaction are considered, with the Coulomb interaction explicitly excluded. As can be seen, the enhancements near the Fermi momentum remain evident, suggesting that this phenomenon is a general feature rather than being limited to specific channels.

\begin{figure}[htbp]
  \centering
  \makebox[\textwidth][c]{\includegraphics[width=0.5\textwidth]{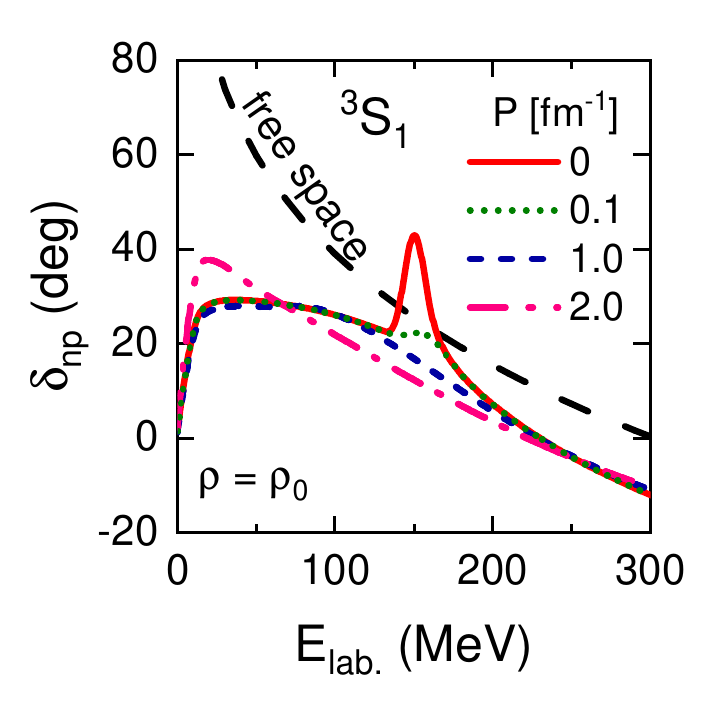}}
  \caption{The $np$ scattering phase shifts for the $^{3}S_{1}$ channel as functions of the laboratory energy. The red solid, green short-dotted, blue short-dashed, and pink short-dash-dotted lines correspond to the total momenta $P=0, 0.1, 1.0$, and $2.0\ \text{fm}^{-1}$, respectively. The phase shifts in free space are also given as black dashed lines for comparison. The density is fixed at $\rho=\rho_0$.}
  \label{Fig2}
\end{figure}

In Fig.~\ref{Fig2}, the in-medium $np$ phase shifts for the $^3S_1$ channel are shown for various total momenta, $P = 0, 0.1, 1.0$, and $2.0\ \text{fm}^{-1}$. It is evident that the enhancement in the phase shift diminishes rapidly for finite total momenta, as demonstrated by the green short-dotted curve corresponding to $P = 0.1\ \text{fm}^{-1}$. This suggests that the peak is closely associated with the Pauli operator $Q_\text{av}(k,P)$, where $P$ plays a critical role. At $P=0$, the Pauli operator acts as a step function with values of either 0 or 1, leading to a sudden change in the integral for $k$ in Eq.~\eqref{eq:ThomEqu} at $k=k_\text{F}$. In contrast, for finite $P$, the Pauli operator $Q_\text{av}(k,P)$ is smoothed, resulting in a more gradual variation in the integral in Eq.~\eqref{eq:ThomEqu}. This provides a technical explanation of the observed enhancement in phase shifts.

On the other hand, the enhancement of the phase shift near the Fermi surface indicates that the two-body interaction is close to a resonant condition, favoring bound or quasi-bound states. In many-body nuclear systems, such near-resonant scattering in the presence of a sharp Fermi surface naturally gives rise to precursor pairing correlations~\cite{2003-Dean-RevModPhys.75.607}. Calculations of pairing gaps using phase-shift–equivalent interactions further show that short-range correlations and resonant scattering strongly influence pairing tendencies~\cite{2017-Rios-JLowTP}. Similarly, a finite temperature smears the Pauli operator and reduces the pairing correlations, much like the effect of a finite total pair momentum $P$. This behavior has been confirmed by previous studies~\cite{2004-Sandulescu-PhysRevC.70.025801}, which show that increasing temperature leads to a decrease of pairing gaps in nuclear matter and neutron star crusts.

Taking a comprehensive view of the figures discussed above, if one disregards the enhancement near the Fermi surface for $P=0$, a general suppression of phase shifts in the medium relative to their free-space values can be observed. For $^1S_0$ and $^3S_1$ channels, this suppression leads to an earlier transition from attractive to repulsive interactions, as indicated by the shift from positive to negative values. For the $^3P_1$ channel, the in-medium phase shifts exhibit larger absolute values. Although the medium effects vary across partial waves, they become more pronounced at higher densities. On the other hand, the phase shifts show limited sensitivity to the total momenta $P$, particularly at higher laboratory energies, as seen in Fig.~\ref{Fig2}. This suggests that the influence of $P$ on the in-medium NN interaction becomes less significant at higher energies.

\begin{figure}[htbp]
  \centering
  \makebox[\textwidth][c]{\includegraphics[width=1.0\textwidth]{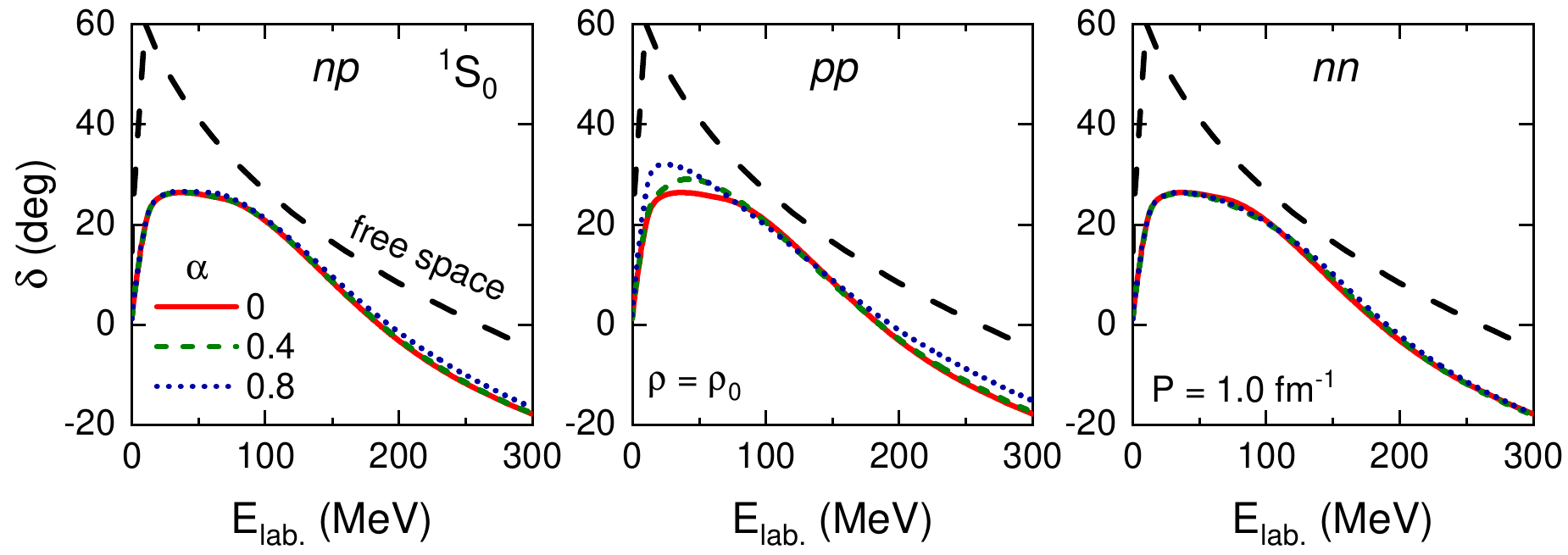}}
  \caption{The $np$, $pp$, and $nn$ scattering phase shifts for the $^{1}S_{0}$ channels as functions of the laboratory energy. In each panel, the red solid, green short-dashed, and blue short-dotted lines represent the results at isospin asymmetries $\alpha = 0, 0.4$, and $0.8$, respectively. The density and total momentum are fixed at $\rho = \rho_0$ and $P=1.0\ \text{fm}^{-1}$. The phase shifts in free space are also given as black dashed lines for comparison.}
  \label{Fig3}
\end{figure}

The preceding analysis has focused on symmetric nuclear matter. In Fig.~\ref{Fig3}, we present the $np$, $pp$, and $nn$ scattering phase shifts for the $^{1}S_{0}$ channel as functions of the laboratory energy, with isospin asymmetries $\alpha = 0, 0.4$, and $0.8$. Notably, the phase shifts exhibit minimal dependence on isospin asymmetry. For the $nn$ pair, the phase shifts can be regarded as isospin independent. A slight variation in phase shifts with $\alpha$ is observed for the $pp$ pair at $E_\text{lab.}\approx 50$\ MeV. Through comprehensive examination of all three subfigures, the theoretically calculated phase shifts for $np$ pairs exhibit markedly stronger alignment with $nn$ results than with $pp$ ones. This demonstrates the neutron dominance in isospin asymmetric nuclear matter.

\subsection{Differential cross sections in nuclear medium}

\begin{figure}[htbp]
  \centering
  \makebox[\textwidth][c]{\includegraphics[width=0.75\textwidth]{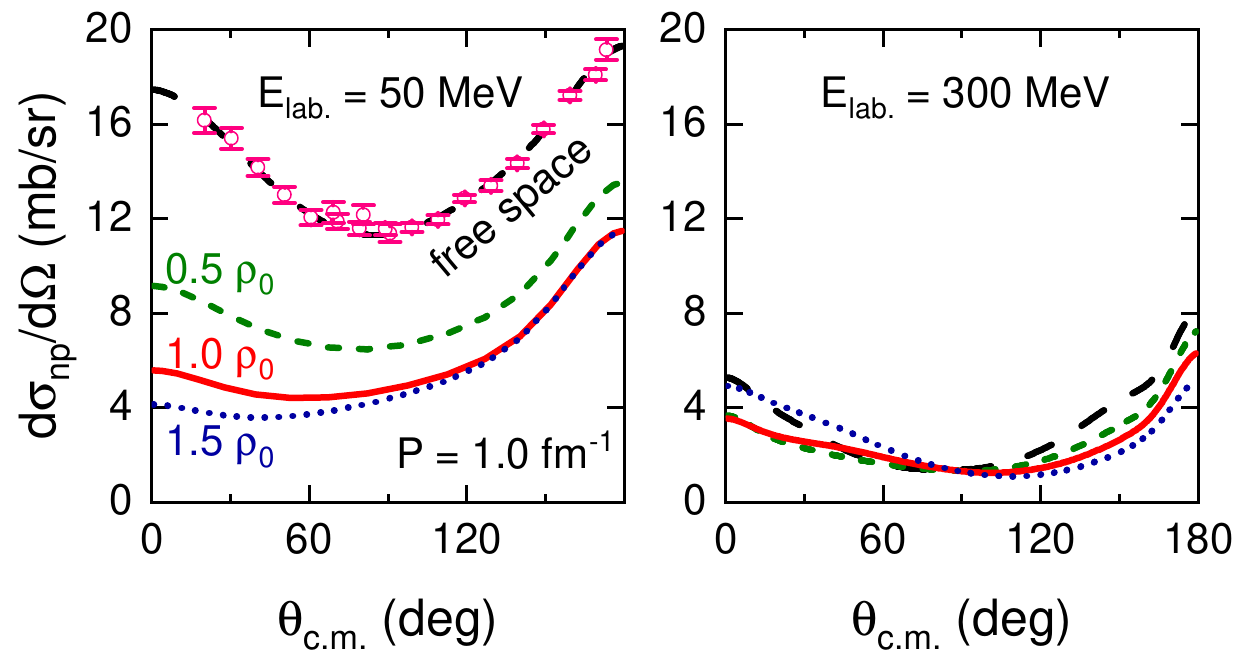}}
  \caption{The $np$ differential scattering cross sections in symmetric nuclear matter as functions of scattering angle at laboratory energy $E_\text{lab.}=50$ and $300$ MeV. Each panel displays results for baryon densities $\rho=0.5\rho_0, \rho_0$, and $1.5\rho_0$, represented by green short-dashed, red solid, and blue short-dotted curves, respectively. The total momentum is fixed at $P=1.0\ \text{fm}^{-1}$. For comparison, free-space phase shifts are included as black dashed lines, while pink symbols with error bars denote experimental data from Ref.~\cite{1977-Montgomery-PhysRevC.16.499}.}
  \label{Fig4}
\end{figure}

Here and in the following, we present the results for differential and total cross sections, comparing them with the corresponding free-space values where applicable. We note that the free-space cross sections include only elastic scattering contributions up to $E_\text{lab.}\approx 300$\ MeV. To properly account for the unitarity violation of the $S$ matrix in the nuclear medium while excluding inelastic scattering effects, we normalize $S^{JS}_{L'L}$ by its modulus in Eq.~\eqref{eq-S2M}. This approach ensures a consistent comparison between medium and free-space cross sections.

Figure~\ref{Fig4} depicts the in-medium $np$ differential scattering cross sections, $d\sigma_{np}/d\Omega$, as functions of the scattering angle at four laboratory energies, $E_\text{lab.}=50$ and $300$\ MeV, and three baryon densities, $\rho = 0.5\rho_0, \rho_0$, and $1.5\rho_0$. The total momentum and isospin asymmetry are fixed at $P=1.0\ \text{fm}^{-1}$ and $\alpha=0$. In comparison to the free-space results denoted by black dashed curves, the in-medium $np$ differential cross sections are suppressed in most cases. They also display a pronounced parabolic profile dominated by enhanced backward-angle scattering. This backward-angle amplification, which becomes more evident with higher densities, originates from spin-isospin exchange interactions inherent to the $np$ system. As $E_\text{lab.}$ increases, both the magnitudes of $d\sigma_{np}/d\Omega$ and its in-medium suppression gradually diminish. Notably, at $E_\text{lab.}=300$\ MeV, the differential scattering cross sections for $1.5\rho_0$ in the range of $30^\circ\leq \theta_{\text{c.m.}}\leq 60^\circ$ exceed their free-space counterparts, marking a reversal of the suppression trend observed at lower energies.

\begin{figure}[htbp]
  \centering
  \makebox[\textwidth][c]{\includegraphics[width=0.75\textwidth]{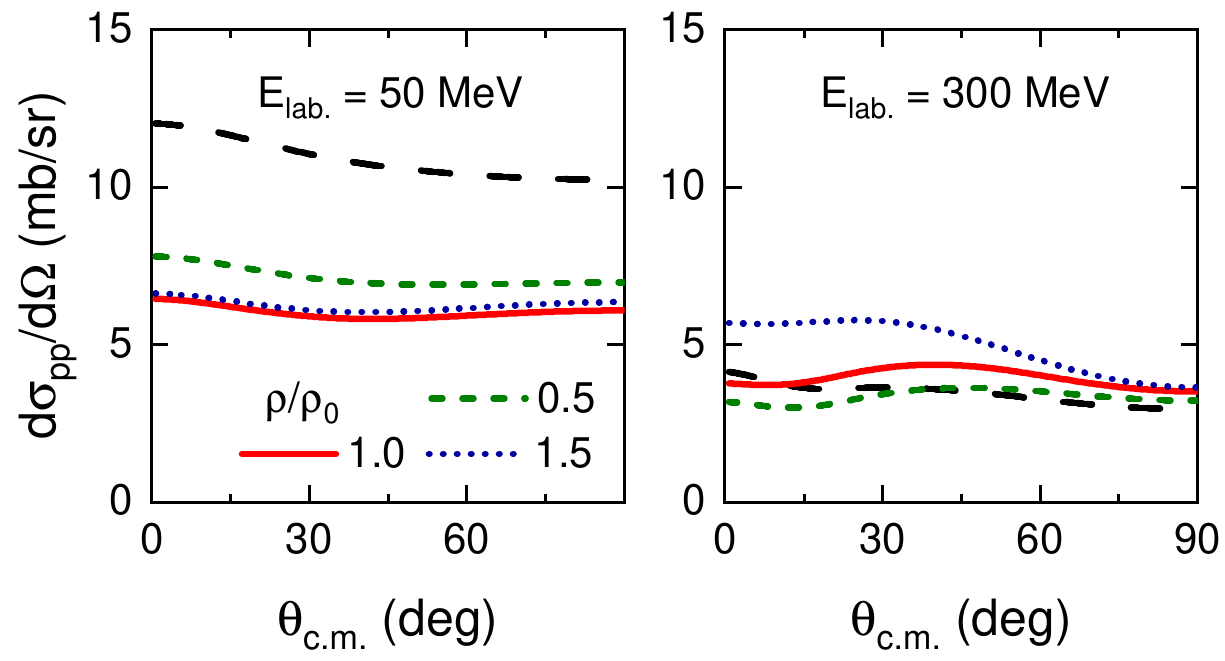}}
  \caption{Same as in Fig.~\ref{Fig4}, but for $pp$ differential cross sections.}
  \label{Fig5}
\end{figure}

Figure~\ref{Fig5} illustrates the $pp$ differential scattering cross sections as functions of the center-of-mass scattering angle for densities $\rho = 0.5\rho_0, \rho_0$, and $1.5\rho_0$, exploring dependencies on $E_\text{lab.}=50$, and $300$\ MeV. Given the symmetric angular distribution of identical fermion pairs scattering with respect to $\theta_\text{c.m.}=90^\circ$, only results for $\theta_\text{c.m.}\leq 90^\circ$ are represented. In contrast to the parabolic angular dependence observed in $np$ scattering (Fig.~\ref{Fig4}), the $pp$ cross sections exhibit a relatively uniform angular distribution. In free space denoted by black dashed curves, $d\sigma/d\Omega$ generally decreases as $E_\text{lab.}$ rises from 50 to 300 MeV. In the nuclear medium, the interplay between energy and density dependencies introduce greater complexity. At $E_\text{lab.}=50$ MeV, $d\sigma/d\Omega$ initially decreases for subsaturation densities $(\rho\leq \rho_0)$ but rises at $\rho=1.5\rho_0$. For $E_\text{lab.}=300$ MeV, $d\sigma/d\Omega$ exhibits a continuous rise across the density range $0.5\rho_0\leq \rho\leq 1.5\rho_0$, ultimately exceeding the corresponding values in free space. An analogous trend for the medium enhancement of $pp$ differential cross sections has been obtained at $E_\text{lab.} = 250$\ MeV in Fig.~3 of Ref.~\cite{CFuchs2001_PRC64-024003}.

\subsection{Cross sections in nuclear medium}

\begin{figure}[htbp]
  \centering
  \makebox[\textwidth][c]{\includegraphics[width=0.75\textwidth]{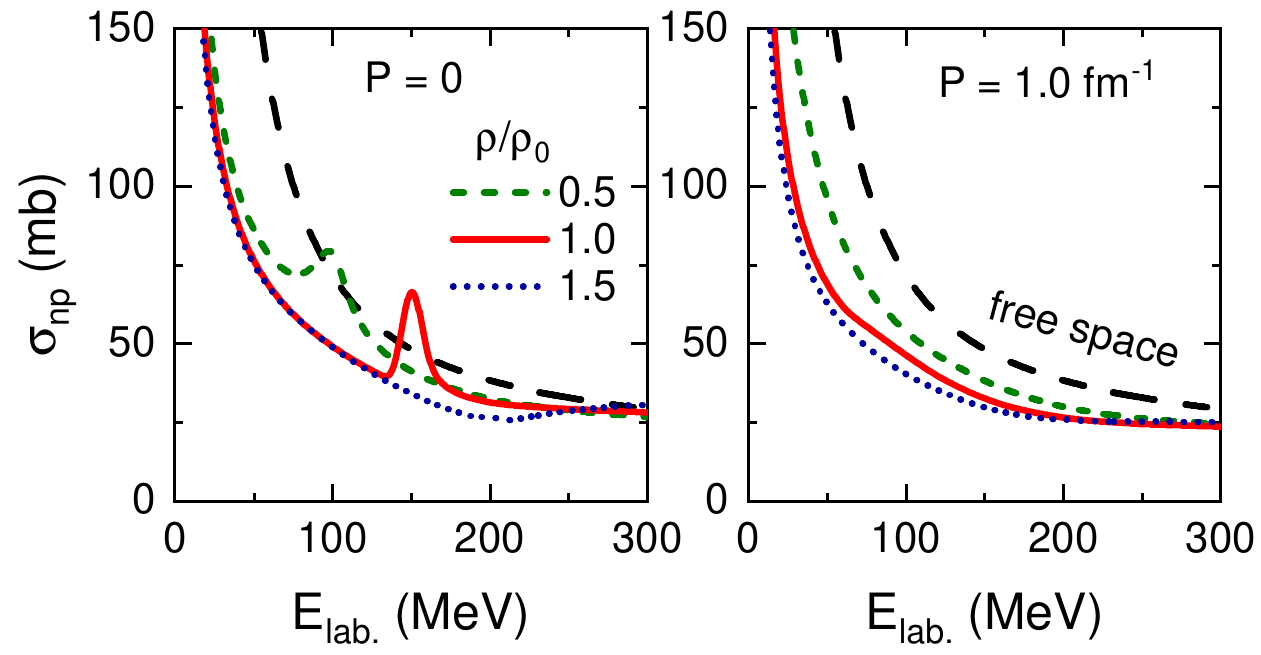}}
  \caption{Total $np$ in-medium cross sections in symmetric nuclear matter as functions of laboratory energy with total momenta $P=0$ and $1.0\ \text{fm}^{-1}$. In each panel, the green short-dashed, red solid, and blue short-dotted curves correspond to the baryon densities $\rho=0, 0.5\rho_0, \rho_0$, and $1.5\rho_0$, respectively. The free-space values are shown as black dashed curves for comparison.}
  \label{Fig6}
\end{figure}

Figure~\ref{Fig6} presents the in-medium total cross sections of $np$ scattering in symmetric nuclear matter, calculated using the RBHF theory in the full Dirac space. The cross sections $\sigma_{np}$ are shown as functions of laboratory energy $E_\text{lab.}$, with variations in total momenta $P = 0$ and $1.0\ \text{fm}^{-1}$ and baryon densities $\rho = 0.08, 0.16, 0.24\ \text{fm}^{-3}$.

A notable characteristic observed in all systems is the sharp decline of $\sigma_{np}$ with increasing $E_\text{lab.}$ below $200$\ MeV, followed by a plateau around 25\ mb at higher energies ($200\leq E_\text{lab.} \leq 300$\ MeV). This rapid decrease at low $E_\text{lab.}$ is consistent with findings from Ref.~\cite{2022-HanSC-PhysRevC.106.064332} using BHF calculations. However, unlike the plateau observed here, the BHF results show monotonic growth at high $E_\text{lab.}$.

Consistent with the phase-shift analysis in Subsection \ref{SubSecI}, significant enhancements in $\sigma_{np}$ are observed near the Fermi surface. In the left panel, the most significant enhancement occurs at $\rho=0.16\ \text{fm}^{-3}$, in qualitative agreement with earlier nonrelativistic BHF calculations \cite{Schulze1997_PRC55-3006}. These enhancements are primarily relevant for total momentum close to zero. Naturally, such enhancements vanish in the latest BHF calculations on in-medium scattering cross sections in Ref.~\cite{2022-HanSC-PhysRevC.106.064332}, where the total momenta of the colliding pairs are constrained to values no less than $0.9\ \text{fm}^{-1}$. Excluding these special enhancements, medium effects generally lead to a reduction in $\sigma_{np}$ compared to the values in free space.

Figure~\ref{Fig6} further reveals that the density dependence of $\sigma_{np}$ is entangled with the total momentum dependence. For scattering in the center-of-mass frame ($P=0$), the total cross sections at different densities largely coincide, except for the aforementioned enhancements. At higher total momenta ($P=1.0\ \text{fm}^{-1}$), deviations between cross sections at varying densities become noticeable, particularly at low laboratory energies.

\begin{figure}[htbp]
  \centering
  \makebox[\textwidth][c]{\includegraphics[width=0.75\textwidth]{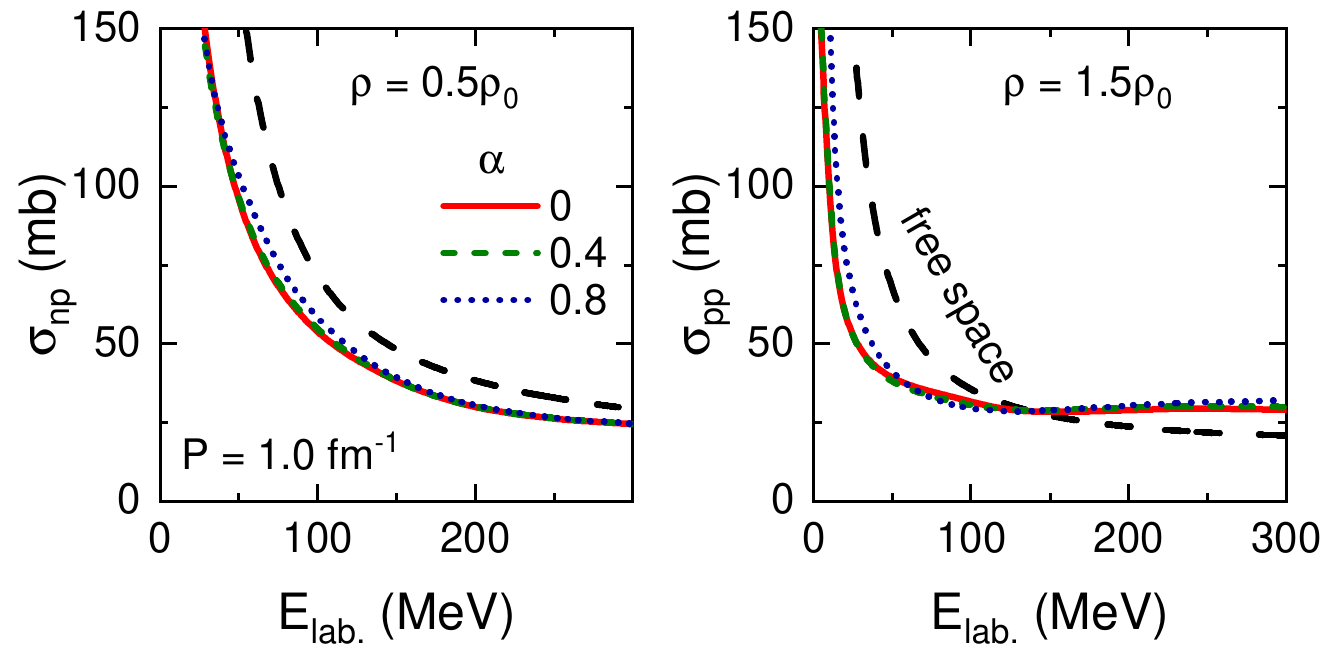}}
  \caption{Total in-medium $np$ cross sections at baryon density $\rho=0.5\rho_0$ and $pp$ cross sections at $\rho=1.5\rho_0$ as functions of laboratory energy. The total momentum is fixed at $P = 1.0\ \text{fm}^{-1}$. In each panel, the red solid, green short-dashed, and blue short-dotted curves correspond to isospin asymmetry $\alpha=0, 0.4$, and $0.8$, respectively. The free-space values are shown as black dashed curves for comparison.}
  \label{Fig7}
\end{figure}

Extending to asymmetric nuclear matter, Fig.~\ref{Fig7} displays the total in-medium $np$ cross sections at $\rho=0.5\rho_0$ and $pp$ cross sections at $\rho=1.5\rho_0$ for isospin asymmetries $\alpha=0, 0.4$, and $0.8$. Across the transition from sub-saturation to supra-saturation densities, the cross sections in symmetric and asymmetric nuclear matter exhibit close agreement, revealing a remarkably weak dependence on isospin asymmetry. This suggests that the parametrizations of $np$ and $pp$ cross sections can be largely simplified by neglecting the $\alpha$-dependence. Notably, while the medium suppression of $np$ scattering continues up to $E_\text{lab.}=300$\ MeV, the $pp$ total scattering cross sections exceed their free-space values at $E_\text{lab.}\approx 150$\ MeV for $1.5\rho_0$. This excess supports previous findings from transport-model investigations of stopping power and collective flow \cite{2006-ZhangYX-PhysRevC.74.014602, 2007-ZhangYX-PhysRevC.75.034615} employing phenomenological in-medium cross sections parameterizations of the form $\sigma_{\text{NN}}=(1-\eta\cdot \rho/\rho_0)\sigma^{\text{free}}_{\text{NN}}$ with $\eta$ being the adjustable factor, as well as studies derived from the effective Lagrangian of density-dependent relativistic hadron theory \cite{2000-LiQF-PhysRevC.62.014606}. However, we note that the transition energy between suppression and excess regimes shows discrepancies among these different approaches.

\begin{figure}[htbp]
  \centering
  \makebox[\textwidth][c]{\includegraphics[width=1.0\textwidth]{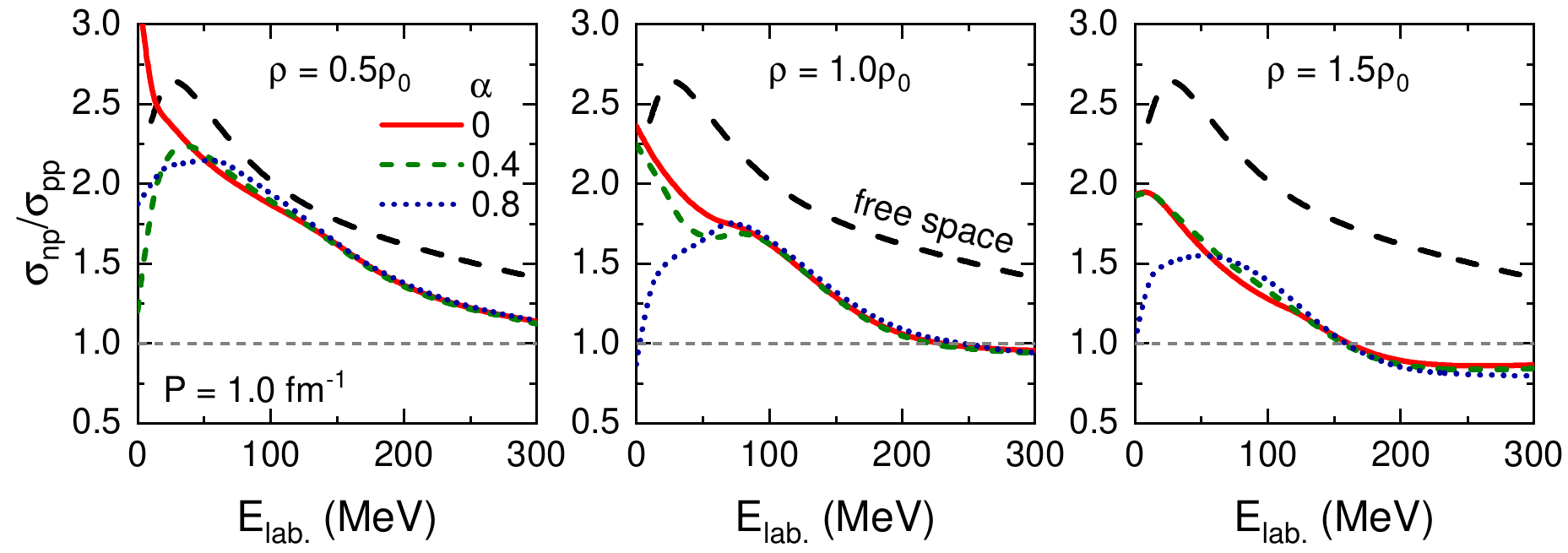}}
  \caption{Ratio of $np$ and $pp$ total scattering cross sections with baryon densities $\rho=0.5, \rho_0$, and $1.5\rho_0$ as functions of laboratory energy. In each panel, the red solid, green short-dashed, and blue short-dotted curves correspond to isospin asymmetry $\alpha=0, 0.4$, and $0.8$, respectively. The total momentum is fixed at $P=1.0\ \text{fm}^{-1}$. The free-space values are shown as black dashed curves for comparison.}
  \label{Fig8}
\end{figure}

To further explore the differences in $np$ and $pp$ cross sections in asymmetric nuclear matter, in Fig.~\ref{Fig8}, we present the ratios of total in-medium cross sections, $\sigma_{np}/\sigma_{pp}$. The free-space ratio deviates from unity, as already known that up to 100 MeV, the free $np$ cross section is about 2-3 times larger than that of $pp$ (and $nn$) scattering \cite{1968-ChenK-PhysRev.166.949}.
This can be attributed to the fact that $np$ scattering involves contributions from both isospin $T=0$ and $T=1$ channels, whereas $pp$ (and $nn$) scattering is restricted to $T=1$ interactions. Our studies show that at low densities $\sigma_{np}$ is larger than $\sigma_{pp}$ up to 300 MeV, and the ratio $\sigma_{np}/\sigma_{pp}$ in medium is mostly influenced by variations in laboratory energy. As $E_\text{lab.}$ increases, the ratio approaches unity and crosses it at $E_\text{lab.}=150$\ MeV for supra-saturation densities. For $E_\text{lab.}\leq 50$\ MeV, the ratio shows irregular behavior with respect to variation of $\alpha$, which is mainly related to the fast change of proton density, as already found in the $^1S_0$ phase shifts of $pp$ in the middle panel of Fig.~\ref{Fig3}. Above 50 MeV, the ratio $\sigma_{np}/\sigma_{pp}$ becomes $\alpha$-insensitive, suggesting simplified future parameterizations.

In the effective mass scaling model, by assuming identical matrix elements for NN interactions in free space and nuclear medium, the in-medium cross sections are expected to be related to the free-space values by a factor of effective masses \cite{Pandharipande1992_PRC45-791, 2002-Persram-PhysRevC.65.064611, LiBA2005_PRC72-064611}, i.e., $\sigma_\text{NN}=(m^*/m)^2\cdot \sigma_\text{NN}^\text{free}$. Our calculations seem to disfavor this simple model, since larger values of $pp$ cross sections as shown in Fig.~\ref{Fig8} is in conflict with $m^*_n>m^*_p$ in asymmetric nuclear matter from our previous studies \cite{2023-WangSB-PhysRevC.108.L031303}. The inadequacy of effective mass scaling model can be attributed to its neglect for the modification of NN interaction matrix elements from the free space to nuclear medium. The modification is easily understood in a relativistic framework where both nuclear force operators and nucleon wave functions are density dependent \cite{Serot1986_ANP16-1}.

\subsection{Benchmarking RBHF methods: NES impact on in-medium cross sections}

\begin{figure}[htbp]
  \centering
  \makebox[\textwidth][c]{\includegraphics[width=0.5\textwidth]{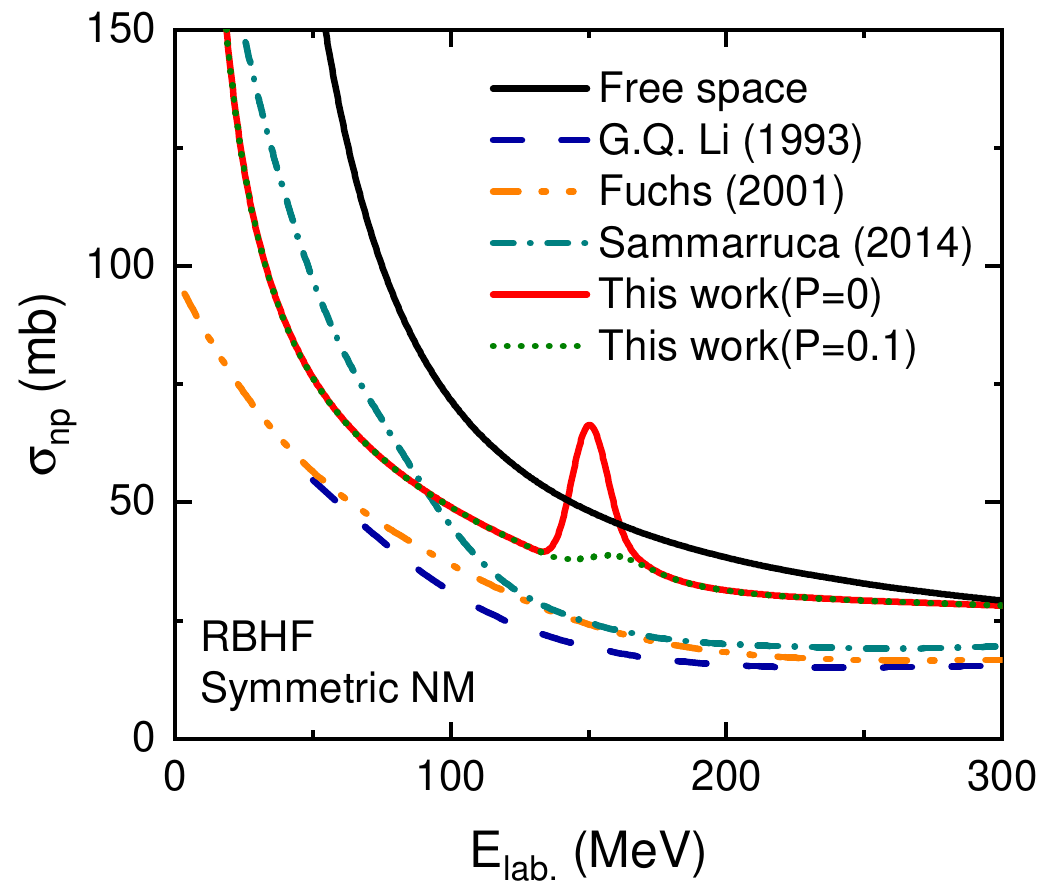}}
  \caption{In-medium $np$ scattering total cross sections obtained in the full Dirac space with $P =0$ and $0.1\ \text{fm}^{-1}$ as functions of laboratory energy, in comparison with the results obtained with PES-only RBHF \cite{GQLi1993_PRC48-1702,CFuchs2001_PRC64-024003,2014-Sammarruca-EPJA}, and that in free space.}
  \label{Fig9}
\end{figure}

To demonstrate the necessity of the full-Dirac-space implementation in RBHF theory, Fig.~\ref{Fig9} compares our calculated $np$ cross sections $\sigma_{np}$ with earlier RBHF results restricted to positive-energy states (PESs). These include the momentum-independent approximation from Refs.~\cite{GQLi1993_PRC48-1702,2014-Sammarruca-EPJA} and the projection method from Ref.~\cite{CFuchs2001_PRC64-024003}. All calculations consider symmetric nuclear matter (NM) near saturation density $\rho \approx 0.16\ \mathrm{fm}^{-3}$ with total momentum $P=0$ for the colliding nucleon pair.

Significant method-dependent variations are evident among the compared results. While general scattering formalisms extend beyond RBHF, their predictive accuracy hinges critically on the quality of the input $G$-matrix. First, the enhanced cross sections at the Fermi momentum are observed only in our framework, demonstrating the critical importance of the exact extraction of momentum-dependent single-particle potentials. We also show the results with finite total momentum $P=0.1\ \text{fm}^{-1}$, indicating the near disappearance of this enhancement. Beyond this enhancement, our $\sigma_{np}$ values fall within the range of prior uncertainties at low $E_\mathrm{lab.}$. Crucially, at higher $E_\mathrm{lab.}$, neglecting NESs leads to underestimated cross sections, aligning with larger derivations on binding energies at high densities as well as single-particle potentials at high momentum \cite{SBWang2021_PRC103-054319}.

Furthermore, comparison to free-space cross sections reveals that only the full-Dirac-space results exhibit the physically expected trend: $\sigma_{np}$ approaches free-space values at high energies, reflecting the weakening of in-medium effects.

\section{Summary and prospective}\label{SecIV}

In summary, this study provides a  microscopic investigation of in-medium nucleon-nucleon (NN) scattering cross sections using the relativistic Brueckner-Hartree-Fock (RBHF) theory with the Bonn potential. By solving the RBHF equations in the full Dirac space, namely incorporating both positive- and negative-energy solutions of the Dirac equation, we derive the complex $G$ matrix in nuclear matter and subsequently calculate the $S$ matrix elements, scattering amplitudes, and observables including phase shifts, differential cross sections, and total cross sections for $pp$, $nn$, and $np$ systems. Our results demonstrate fundamental modifications to NN interactions in dense hadronic environments. The absence of bound states in nuclear matter is revealed through characteristic phase-shift patterns in both $^1S_0$ and $^3S_1$ channels, while medium suppression effects exhibit stronger sensitivity to baryon density than to isospin asymmetry. Distinct angular dependencies emerge between $np$ and $pp$ scattering, with the former displaying pronounced backward-angle enhancement and the latter showing quasi-isotropic distributions. Notably, resonant-like amplifications of phase shifts and total cross sections are observed near the Fermi momentum at zero total momentum, correlating with the emergence of pairing correlation in nuclear matter.

The analysis underscores the complicated multi-parameter dependence of in-medium cross sections on laboratory energy $E_\text{lab.}$, total momentum $P$, density $\rho$, and isospin asymmetry $\alpha$. To improve the practical applicability of our results, we are developing a systematic parameterization scheme of $\sigma(E_\text{lab.}, P, \rho, \alpha)$ to facilitate implementation in transport models. This parametrization facilities the mean field calculated self-consistently starting from the same realistic NN interactions \cite{SBWang2022_PRC106-L021305}. Future efforts will extend this framework to include Deltas in the scattering processes \cite{1987-Haar-PhysRevC.36.1611, 2019-LiQF-SciChina, 2024-NanMZ-EPJA}, providing critical elastic and inelastic cross sections essential for modeling intermediate-energy heavy ion collision simulations. The present calculations also provide a first step toward applying RBHF-based in-medium cross sections to astrophysical environments. For realistic applications to neutron star physics, however, one must consider strongly asymmetric nuclear matter as well as higher densities, which will be the subject of future investigations. Additionally, recent years have witnessed important progress in developing relativistic chiral nuclear forces \cite{2025-LuJX-arXiv,2025-ShenSH-arXiv}. Microscopic in-medium cross sections starting from that would be more promising considering its closer relation to the chiral symmetry and its breaking in low-energy quantum chromodynamics.

\begin{acknowledgments}

This work was partly supported by the Guizhou Provincial Science and Technology Projects under grant No. ZK[2022]203, PhD fund of Guizhou Minzu University under grant No. GZMUZK[2024]QD76, the National Natural Science Foundation of China under grant Nos. 12205030, 12265012, 12405143, and the Project No. 2024CDJXY022 supported by the Fundamental Research Funds for the Central Universities. Part of this work was achieved by using the supercomputer SQUID at the Cybermedia Center, Osaka University under the support of Research Center for Nuclear Physics of Osaka University. Part supported by the European Research Council (ERC) under the European Union's Horizon 2020 research and innovation program (AdG EXOTIC, grant agreement No. 101018170) and by the MKW NRW under the funding code NW21-024-A.

\end{acknowledgments}

\bibliography{ImSigma}

\end{document}